\documentclass[runningheads]{llncs}
\usepackage[T1]{fontenc}
\usepackage{graphicx}
\usepackage{subfig}         
\usepackage{pgf}            
\graphicspath{{figures/}}  
\usepackage{amsmath}
\usepackage{amssymb}
\usepackage{mathtools}
\usepackage{bbm}
\usepackage{hyperref}
\usepackage{color}

\usepackage{algorithm}        
\usepackage{algpseudocodex}   
\DeclareMathOperator*{\argmin}{arg\,min}
\DeclareMathOperator*{\argmax}{arg\,max}

\newcommand{\mat}[1]{\mathbf{#1}}   

\newcommand{\tran}{^{\mathsf{T}}}
\DeclareMathOperator{\E}{\mathbb{E}}        

\newcommand{\C}{\mathbb{C}}

\DeclarePairedDelimiter{\abs}{\lvert}{\rvert}
\DeclarePairedDelimiter{\norm}{\lVert}{\rVert}

\newcommand{\given}{\mathbin{\vert}}        
\begin{document}
\title{Deep Shape Regression for Planar Curves\\ with Multimodal Covariates}
\author{Manuel Pfeuffer\inst{1}\thanks{Corresponding author.} \and
Roshan P.\ Rane\inst{1,2,3} \and
Hadya Yassin\inst{4} \and
Kerstin Ritter\inst{2,3} \and
Sonja Greven\inst{1}
for the Alzheimer's Disease Neuroimaging Initiative\thanks{Data used in preparation of this article were obtained from the Alzheimer's Disease Neuroimaging Initiative (ADNI) database (\url{adni.loni.usc.edu}). As such, the investigators within the ADNI contributed to the design and implementation of ADNI and/or provided data but did not participate in analysis or writing of this report. A complete listing of ADNI investigators can be found at: \url{http://adni.loni.usc.edu/wp-content/uploads/how_to_apply/ADNI_Acknowledgement_List.pdf}}}
\authorrunning{M. Pfeuffer et al.}
\institute{Humboldt-Universität zu Berlin, Berlin, Germany\\ \email{manuel.pfeuffer@hu-berlin.de} \and
Hertie Institute for AI in Brain Health, Universität Tübingen, Tübingen, Germany \and
Universität Tübingen, Tübingen, Germany \and
Hasso Plattner Institut, Universität Potsdam, Potsdam, Germany}
\maketitle              
\begin{abstract}
The shape of a planar curve is the geometric information that remains once translation, rotation, scale and reparametrisation are removed and is of interest in many health applications, e.g. in neuroimaging.
We propose a deep shape regression model for open planar curves that admits multimodal and
high-dimensional covariates. 
Representing curves as complex-valued functions, we show that the conditional full Procrustes mean is the leading eigenfunction of the conditional covariance.
To estimate this covariance surface, we propose a novel deep conditional covariance smoother with modality-specific encoders --- e.g.\ splines for scalar covariates and convolutional networks for images, which classical spline smoothers cannot accommodate.
Our model is by construction invariant to translation, rotation and scaling of the input curves and handles sparsely and irregularly sampled curves.
We further provide an algorithm for elastic mean estimation that also removes parametrisation by iterating covariance smoothing, rotational alignment and parametrisation alignment.
We illustrate the method on simulated outlines with known conditional mean and multimodal covariates, and give a first application to hippocampal outlines from the ADNI cohort, recovering covariate effects consistent with the literature.
Code is available at \url{https://github.com/mpff/dnn-shapes}.
\end{abstract}
\section{Introduction}
\label{sec:intro}
In statistical shape analysis, the shape of a planar curve $y : [0,1] \to \mathbb{R}^2$ is its equivalence class under translation, rotation, scaling and reparametrisation \cite{dryden2016,srivastava2016}.
For instance, the outline of a bone or organ traced from a medical image may be positioned differently across observations purely because of differences in patient positioning or image acquisition, which carries no shape information.
In this paper, we aim to estimate how the shape of an outline varies with covariates $\vec x$ that may be high-dimensional and multimodal, combining for instance tabular demographics, disease status, health records, genomics and imaging.
We propose a novel deep shape regression model for planar curves, which combines methods from several research fields: 
\emph{Statistical shape analysis} supplies the notion of a full Procrustes mean, a popular mean concept that is invariant under translation, scaling and rotation of the observations and has convenient estimation properties for two-dimensional shapes \cite{dryden2016}.
\emph{Elastic functional data analysis} represents the outline as a continuous curve rather than a fixed set of landmark points.
It retains the geometric information along the whole object and then additionally considers the invariance with respect to reparametrisation of curves \cite{srivastava2011ElasticCurves}.
Finally, \emph{deep learning} provides the encoders and gradient-descent-based fitting routine that let high-dimensional and multimodal covariates enter the model.

Our approach builds on the result that the full Procrustes mean of a sample of planar curves is the leading eigenfunction of their Hermitian covariance \cite{stoecker2026elasticfullproc}, which can be derived when treating a planar curve as a complex-valued function $y : [0,1] \to \mathbb{C}$.
We extend this in two ways.
First, we show that the same characterisation holds conditionally: the \emph{conditional full Procrustes mean} is the leading eigenfunction of the \emph{conditional covariance} $C(s,t \given \vec x) = \E[y(s)\,\overline{y(t)} \given \vec x]$, where $\overline{z}$ denotes the complex conjugate.
Second, to estimate the conditional mean, we propose a novel \emph{deep conditional covariance smoother}, with modality-specific encoders, to carry out this smoothing with covariates that a classical tensor-product spline smoother \cite{ding2022FunctionalPCA} cannot accommodate.
We demonstrate the method on conditional hippocampus shape-and-scale mean estimation, recovering the effect of each covariate on the outline.

The full Procrustes mean is invariant to translation, rotation and scaling, but treating the outline as a continuous curve rather than a set of landmarks adds a second necessary invariance: reparametrisation, the (arbitrary) speed at which the curve is traced \cite{srivastava2016}.
Ignoring it leaves the estimated mean visibly washed out at sharp features due to misalignment of different parts of the curve across observations, as can be seen in our simulation study (Section~\ref{sec:simulation}).
A reparametrisation-invariant mean is called \emph{elastic} and it is usually estimated by parametrisation alignment of the observed curves \cite{steyer2023ElasticAnalysis}.
However, as this is not rotation-invariant, rotation and reparametrisation must be aligned separately.
We extend \cite{stoecker2026elasticfullproc} and propose an estimation algorithm for \emph{conditional elastic full Procrustes means} that iterates covariance smoothing, rotational alignment, and parametrisation alignment.
St\"ocker et al.~\cite{stoecker2026elasticfullproc} and Srivastava et al.~\cite{srivastava2011ElasticCurves} estimate elastic shape means, but only unconditionally.
Covariate effects on functional outlines have otherwise been modelled by St\"ocker et al.~\cite{stocker2023FunctionalAdditive} who fit additive models on manifolds of planar shapes but do not consider reparametrisation, and Steyer et al.~\cite{steyer2026} who developed model-based Fréchet regression in quotient metric spaces for elastic curves but are not rotationally invariant.
Both predict a conditional mean outline under some of our invariances but consider only tabular covariates, whereas our encoders also admit images and other modalities.

The deep conditional covariance smoother we propose is a general method for conditional covariance estimation in its own right, making it broadly applicable in functional data analysis.
Covariance smoothing typically estimates a single, unconditional covariance \cite{yao2005FunctionalData}, but conditional extensions exist.
Ding et al.~\cite{ding2022FunctionalPCA} let both the functional mean and the covariance depend on covariates through spline bases and roughness penalties, but their construction is restricted to scalar covariates.
Neural models have been used for estimating the covariance: CovNet \cite{sarkar2022CovNet} provides a flexible covariance network with a closed-form eigendecomposition, yet it is unconditional.
A separate line of work replaces the covariance by a neural encoder of the curves: Zhong et al.~\cite{zhong2026Nonlinear} perform nonlinear functional principal component analysis aimed at curve reconstruction, and Wu et al.~\cite{wu2023NeuralNetworks}, Yao et al.~\cite{yao2021AdaptiveBasis} and Rao and Reimherr~\cite{rao2023NonlinearFunctional} model the conditional mean function from scalar inputs through learned bases.
In this work we adapt the approach by Ding et al.~\cite{ding2022FunctionalPCA} to Hermitian covariance smoothing and to admit deep encoders instead of spline bases of potentially multimodal covariates, thereby bridging the gap between neural functional models and conditional covariance smoothing.

\section{Method}
\label{sec:method}

Identifying $\mathbb{R}^2$ with $\mathbb{C}$, a planar curve becomes a complex-valued function $y : [0,1] \to \mathbb{C}$, canonically parametrised over the unit interval.
We write $\overline{z}$ for the complex conjugate and $|z| = \sqrt{z\,\overline z}$ for the modulus.
On $\mathbb{L}^2([0,1],\mathbb{C})$ we use the Hermitian inner product $\langle f, g\rangle = \int_0^1 f(t)\,\overline{g(t)}\,\mathrm dt$ and the induced norm $\norm{f} = \sqrt{\langle f, f\rangle}$.
Rotation by an angle $\theta$ then acts as $y \mapsto e^{i\theta} y$, scaling by a factor $\lambda > 0$ as $y \mapsto \lambda y$, and translation as $y \mapsto y + \xi$ for $\xi \in \mathbb{C}$.
A joint rotation and scaling is therefore multiplication by a single complex number $\omega = \lambda e^{i\theta} \in \mathbb{C}$, which renders the two-dimensional case analytically convenient \cite{dryden2016}.
Since only the image of each curve is observed and not its parametrisation, we fix an initial parametrisation as described in Section~\ref{sec:setup} and defer the reparametrisation-invariant elastic full Procrustes mean to Section~\ref{sec:elastic}.

\subsection{Pre-shape and Arc-length Representation}
\label{sec:setup}
Consider a sample of pairs $(\vec x_i, \vec y_i)$, $i = 1, \dots, n$, where the $J$ covariates $\vec x_i = (x_{i1}, \dots, x_{iJ})$ are possibly multimodal and $\vec y_i = (y_{i1}, \dots y_{im_i})$ are the $m_i$ irregularly and (possibly) sparsely observed points along the response object $y_i(t): [0,1] \to \mathbb{C}$, an open planar curve.
Note that we don't assume that the $y_i(t)$'s are sampled on a common grid or even with the same number of observations per curve.
As an example, $\vec x_i$ could be a collection of observed demographic variables, \emph{Alzheimer's disease} (AD) status, gene expressions and a \emph{structural magnetic resonance image} (sMRI), while $\vec y_i$ could be a collection of points along the outline of the hippocampus.
As we only observe the 2D coordinates $y_{ij} = y_i(t_{ij})$, but not their parametrisation $t_{i1} < \dots < t_{in_i} \in [0,1]$, we estimate a starting parametrisation for curve $i$ by using the observed \emph{arc-length} up to point $j$, $t_{ij} = \sum_{l=2}^{j}\abs{y_{il}-y_{i,l-1}} / \sum_{l=2}^{m_i}\abs{y_{il}-y_{i,l-1}}$, with $t_{i1}=0$ and $t_{im_i}=1$.
Note that this does not guarantee exact alignment of observed points with respect to parametrisation, and to obtain better shape mean estimates, we should also include the invariance with respect to reparametrisation (see Section~\ref{sec:elastic}).

After obtaining a parametrisation, we remove translation and scaling of each curve as a pre-processing step.
Translation is removed by centring each curve using its weighted centroid $\bar y_i = \sum_j w_{ij}\, y_{ij}$, with trapezoidal weights $w_{ij} = (t_{i,j+1}-t_{i,j-1})/2$, accounting for the irregular spacing of observed points, and half-intervals at the endpoints.
Scaling is removed by rescaling to unit norm $\norm{y_i - \bar y_i}_w^2 = \sum_j w_{ij}\,\abs{y_{ij}-\bar y_i}^2$, leading to centred, unit-norm curves $\tilde y_{ij} = (y_{ij}-\bar y_i)/\norm{y_i - \bar y_i}_w$.
The $\tilde y_i$ are a \emph{pre-shape} representation of $y_i$, with translation and scaling removed, but before filtering out rotational effects.

\subsection{The Full Procrustes Distance}
\label{sec:met:fpdist}
For two curves with centred and unit-norm representatives $\tilde y_1, \tilde y_2$, a notion of the difference in their shape is given by the \emph{full Procrustes distance}, which optimises over rotation and scaling, translation having already been removed by centring,
\begin{equation}
  d(\tilde y_1, \tilde y_2) = \min_{\omega \in \mathbb{C}} \norm{\tilde y_1 - \omega\tilde y_2}.
  \label{eq:procdist}
\end{equation}
Conveniently, this admits the closed form $d(\tilde y_1, \tilde y_2) = \sqrt{\,1 - \langle \tilde y_1, \tilde y_2\rangle\langle \tilde y_2, \tilde y_1\rangle\,}$, with the minimum attained at $\omega = \langle \tilde y_1, \tilde y_2\rangle$ \cite{dryden2016,stoecker2026elasticfullproc}
(see the Appendix).
Using this representation, St\"ocker et al.~\cite{stoecker2026elasticfullproc} show that the \emph{full Procrustes mean} of a sample of curves is the leading eigenfunction of their Hermitian covariance surface $C(s,t) = \E[\tilde y(s)\,\overline{\tilde y(t)}]$, satisfying $C(s,t) = \overline{C(t,s)}$, turning mean estimation into a covariance smoothing problem.
Note that the full Procrustes distance optimises over scaling again, even though we use rescaled pre-shapes as inputs.
This makes it more robust towards outliers (that can be shrunk towards zero), especially in mean estimation.
\cite{dryden2016} contains a detailed explanation of this together with a comparison to other notions of shape distance.
In this paper we choose to use the full Procrustes distance mainly because of its convenient connection to covariance smoothing when estimating the full Procrustes mean.
This is derived in the next section.

\subsection{The Functional Conditional Full Procrustes Mean}
\label{sec:eigen}
We now make precise the shape mean we estimate and show that it is the leading eigenfunction of the conditional covariance of the pre-shape curves.
Let $\tilde y$ be the random pre-shape curve of Section~\ref{sec:setup} and let us define the \emph{conditional full Procrustes mean} as the minimiser of the expected squared full Procrustes distance to $\tilde y$, over all unit-norm $\mathbb{L}^2$-curves, conditional on covariate value $\vec x$.
Then
\begin{align}
  \mu(t \given \vec x)
  = & \argmin_{\mu(\vec x) \in \mathbb{L}^2,\, \norm{\mu(\vec x)}=1}\
    \E\!\left[\, d(\mu(\vec x), \tilde y)^2 \,\given\, \vec X = \vec x \,\right]\\
  = & \argmax_{\mu(\vec x) \in \mathbb{L}^2,\, \norm{\mu(\vec x)}=1}\
    \E\!\left[\, \langle \mu(\vec x), \tilde y\rangle \langle \tilde y, \mu(\vec x) \rangle \,\given\, \vec X = \vec x \,\right], \label{eq:fp-argmax}
\end{align}
where the second line follows from the closed form full Procrustes distance and dropping the constant.
Writing out the complex functional scalar products and exchanging conditional expectation and integration, we can rewrite~\eqref{eq:fp-argmax} as
\begin{align}
  \mu(t \given \vec x)
  = & \argmax_{\mu(\vec x) \in \mathbb{L}^2,\, \norm{\mu(\vec x)}=1}\
    \E\!\left[\, \iint_0^1 \mu(s \given \vec x) \overline{\tilde y(s)} \tilde y(t) \overline{\mu(t \given \vec x)} \mathrm{d}s \mathrm{d}t \,\given\, \vec X = \vec x \,\right]\\
  = & \argmax_{\mu(\vec x) \in \mathbb{L}^2,\, \norm{\mu(\vec x)}=1}\
    \iint_0^1 \mu(s \given \vec x) \E\!\left[\, \tilde y(s)\, \overline{\tilde y(t)} \,\given\, \vec X = \vec x \,\right] \overline{\mu(t \given \vec x)} \mathrm{d}s \mathrm{d}t.
\end{align}
We can see that the optimization fully depends on the \emph{conditional covariance surface}
\begin{equation}
  C(s,t \given \vec x) = \E\!\left[\, \tilde y(s)\, \overline{\tilde y(t)} \,\given\, \vec X = \vec x \,\right],
  \qquad C(s,t \given \vec x) = \overline{C(t,s \given \vec x)} ,
  \label{eq:ccov}
\end{equation}
which is Hermitian and positive semi-definite for every $\vec x$.
We can rewrite the optimization problem in operator form as
\begin{equation}
  \mu(t \given \vec x) = \argmax_{\mu(\vec x) \in \mathbb{L}^2,\, \norm{\mu(\vec x)}=1}\ \langle \mu(\vec x),\, \mathcal{C}_{\vec x}\,\mu(\vec x)\rangle ,
\end{equation}
with $(\mathcal{C}_{\vec x} f)(s) = \int_0^1 C(s,t \given \vec x)\, f(t)\, \mathrm{d}t$.
Maximizing $\langle \mu(\vec x), \mathcal{C}_{\vec x}\,\mu(\vec x)\rangle$ over the self-adjoint, positive semi-definite operator $\mathcal{C}_{\vec x}$ at $\norm{\mu(\vec x)} = 1$ is a classical functional principal component analysis (FPCA) problem \cite{ramsay2005functional,hsing2015theoretical}.
Let $\mathcal{C}_{\vec x}$ have eigenpairs $\mathcal{C}_{\vec x}\, u_r(t, \vec x) = \lambda_r(\vec x)\, u_r(t, \vec x)$ with $\lambda_1(\vec x) \ge \lambda_2(\vec x) \ge \cdots \ge 0$. 
The conditional full Procrustes mean at $\vec x$ is the normalized leading eigenfunction
\begin{equation}
  \mu(t,\vec x) = u_1(t,\vec x),
  \label{eq:eigfun}
\end{equation}
determined up to a rotation and maximizing $\lambda_1(\vec x) = \langle u_1(\vec x), \mathcal{C}_{\vec x} u_1(\vec x) \rangle$.
The trailing eigenfunctions $u_2(t, \vec x), u_3(t, \vec x), \dots$ are the conditional modes of shape variation, each explaining a share $\lambda_r(\vec x)$ of the variation. 

\subsection{Deep Conditional Covariance Smoothing}
\label{sec:dccs}

We model the conditional covariance surface~\eqref{eq:ccov} through a cubic B-spline basis $\vec b(t) \in \mathbb{R}^{k}$ with $k-4$ interior knots over $t \in [0,1]$, writing
\begin{equation}
  \hat C(s,t \given \vec x) = \vec b(s)\tran \mat\Xi(\vec x)\, \vec b(t),
  \qquad \mat\Xi(\vec x) \in \mathbb{C}^{k \times k}.
  \label{eq:cov-factor}
\end{equation}
The covariate dependence is carried entirely by the Hermitian coefficient matrix $\mat\Xi(\vec x)$.
We follow the approach of Ding et al.~\cite{ding2022FunctionalPCA} and model it as an outer product
\begin{equation}
  \mat\Xi(\vec x) = \vec\theta(\vec x)\, \vec\theta(\vec x)^{\mathsf H},
  \qquad \vec\theta(\vec x) \in \mathbb{C}^{k \times r},
  \label{eq:rank1}
\end{equation}
where $(\cdot)^{\mathsf H}$ denotes the conjugate transpose and $r \le k$ can be chosen to control the rank of $\mat \Xi (\vec x)$.
The outer product makes $\mat\Xi(\vec x)$ Hermitian and positive semi-definite for any $\vec\theta(\vec x)$ with no constraints on the entries of $\vec\theta(\vec x)$.
We give $\vec\theta(\vec x)$ an additive structure over the $J$ covariates,
\begin{equation}
  \vec\theta(\vec x) = \vec\theta_0 + \sum_{j=1}^{J} \vec\theta_j(x_j),
  \qquad \vec\theta_0,\, \vec\theta_j(x_j) \in \mathbb{C}^{k \times r},
  \label{eq:theta-additive}
\end{equation}
with a covariate-free intercept $\vec\theta_0$ and one additive effect $\vec\theta_j$ per covariate.
The additivity is a modelling choice and extensions to interactions of covariates would be possible.
Each effect is linear in a covariate-specific (deep) embedding $\vec\varphi_j(x_j) \in \mathbb{R}^{q_j}$,
\begin{equation}
  \vec\theta_j(x_j) = \mat W_j\, \vec\varphi_j(x_j),
  \qquad \mat W_j \in \mathbb{C}^{k \times r \times q_j},
  \label{eq:theta-readout}
\end{equation}
which mirrors the covariate-basis construction of~\cite{ding2022FunctionalPCA}.
The embedding $\vec\varphi_j$ can be chosen per covariate and can be a mix of pre-defined bases and neural encoders:
a spline basis evaluation when $x_j$ is scalar, giving a smooth univariate effect; a linear map for a dummy covariate; a multilayer perceptron for a tabular vector; or a convolutional encoder for an sMRI image or gene sequence.
Because additivity and the embeddings act only through $\vec\theta(\vec x)$, the surface~\eqref{eq:rank1} stays Hermitian and positive semi-definite for arbitrary, unconstrained encoders.

We fit the surface by covariance smoothing from the observed points, the conditional and neural analogue of the bivariate spline smoothing of St\"ocker et al.~\cite{stoecker2026elasticfullproc}, which in turn follows \cite{cederbaum2018fast}.
For curve $i$ observed at $t_{i1}, \dots, t_{im_i}$, the off-diagonal products
\begin{equation}
  c_{i,lm} = \tilde y_i(t_{il})\, \overline{\tilde y_i(t_{im})},
  \qquad l < m,
  \label{eq:rawcov}
\end{equation}
are raw observations of the covariance, $\E[c_{i,lm} \given \vec x_i] = C(t_{il}, t_{im} \given \vec x_i)$.
The diagonal $l = m$ is excluded as it carries the measurement-error variance, which would inflate the fitted surface (see~\cite{yao2005FunctionalData,cederbaum2018fast} for details).
We estimate the parameters of the covariance surface and encoders by penalised least squares,
\begin{equation}
  \min_{\vec\theta_0,\, \{\mat W_j\},\, \{\vec\varphi_j\}}\
  \sum_{i} \sum_{l < m} \bigl|\, c_{i,lm} - \hat C(t_{il}, t_{im} \given \vec x_i) \,\bigr|^2
  \;+\; \mathcal{P},
  \label{eq:objective}
\end{equation}
minimised by stochastic gradient descent over minibatches of curves, with penalty $\mathcal{P}$. 

The penalty controls the roughness of the mean in $t$.
The $t$-dependence of the mean enters only through $\vec b(t)$ with coefficients $\vec\theta(\vec x)$, so a second-order difference penalty on these coefficients directly controls the wiggliness of the estimated mean shape.
We penalise the baseline and each covariate effect,
\begin{equation}
  \mathcal{P}_t = \rho_t \Bigl( \norm{\mat\Delta^2 \vec\theta_0}^2 + \sum_{j=1}^{J} \norm{\mat\Delta^2 \mat W_j}_F^2 \Bigr),
  \label{eq:pen-t}
\end{equation}
where $\mat\Delta^2$ is the second-order difference operator along the spline index and $\rho_t \ge 0$ is a hyperparameter setting the smoothness.
Penalising $\mat W_j$ rather than $\vec\theta_j(x_j)$ enforces the same smoothness for every value of the covariate.
If, instead of a neural network, we also want to specify a spline basis for a scalar covariate, a difference penalty $\mathcal P_x$ along the embedding index may additionally be imposed for a smooth dependence on $x_j$, using hyperparameters $\rho_j \ge 0$.
The total penalty is $\mathcal{P} = \mathcal{P}_t + \mathcal{P}_x$.
The penalty weights $\rho_t$ and $\rho_j$ can be selected by grid search on a held-out validation set.

\subsection{Mean Estimation Using Eigendecomposition}
\label{sec:mean}

The conditional full Procrustes mean is the leading eigenfunction of the conditional covariance operator.
For the fitted surface~\eqref{eq:cov-factor} this eigenproblem is finite-dimensional.
Let $\hat C(s,t \given \vec x) = \vec b(s)\tran \mat\Xi(\vec x)\, \vec b(t)$ with $\mat\Xi(\vec x) \in \mathbb{C}^{k \times k}$ Hermitian positive semi-definite, and write $\mat B = \int_0^1 \vec b(t)\,\vec b(t)\tran\,\mathrm{d}t$ for the real, symmetric positive definite Gram matrix.
A function $f = \vec c\tran \vec b$ with $\vec c \in \mathbb{C}^{k}$ is then an eigenfunction of $\hat{\mathcal C}_{\vec x}$ with eigenvalue $\lambda$ if and only if $\mat\Xi(\vec x)\,\mat B\,\vec c = \lambda\,\vec c$, a generalised eigenproblem, with norm $\norm{f}^2 = \vec c^{\mathsf H}\mat B\,\vec c$ \cite{ramsay2005functional}.
Substituting $\vec d = \mat B^{1/2}\vec c$ turns this into the ordinary Hermitian eigenproblem $\mat B^{1/2}\mat\Xi(\vec x)\,\mat B^{1/2}\,\vec d = \lambda\,\vec d$, whose eigenpairs we order as $\lambda_1(\vec x) \ge \lambda_2(\vec x) \ge \cdots \ge 0$.
The conditional full Procrustes mean is the unit-norm leading eigenfunction
\begin{equation}
  \hat\mu(t \given \vec x) = \vec c_1(\vec x)\tran \vec b(t),
  \qquad \vec c_1(\vec x) = \mat B^{-1/2}\vec d_1(\vec x).
  \label{eq:mean}
\end{equation}
In practice we do not form $\mat\Xi(\vec x) = \vec\theta(\vec x)\vec\theta(\vec x)^{\mathsf H}$ explicitly:
with the singular value decomposition $\mat B^{1/2}\vec\theta(\vec x) = \mat U\mat\Sigma\mat V^{\mathsf H}$ we have $\mat B^{1/2}\mat\Xi(\vec x)\mat B^{1/2} = \mat U\mat\Sigma^2\mat U^{\mathsf H}$, so $\vec d_r(\vec x)$ is the $r$-th column of $\mat U$ and $\lambda_r(\vec x) = \Sigma_{rr}^2$, the $r$-th diagonal element of $\mat\Sigma^2$.
The leading left singular vector of $\mat B^{1/2}\vec\theta(\vec x)$ therefore gives the mean directly, and $\mat B$ is the only quantity that depends on the basis.

\subsection{Elastic Mean Estimation Using Iterative Alignment}
\label{sec:elastic}
The pre-shape representation of Section~\ref{sec:setup} uses an initial arc-length parametrisation. 
It removes translation and scaling, and rotation is removed in the covariance estimation step.
This, however, does not account for \emph{reparametrisation} invariance with respect to warping: applying a warping function $\gamma : [0,1] \to [0,1]$ replaces a curve $y$ by $y \circ \gamma$, which traces the same image at a different speed \cite{srivastava2016}.
Warping is a type of shape invariance that is unique to treating outlines as curves $y(t)$, instead of a collection of landmark points.
Without accounting for it, two outlines traced at different speeds are treated as different shapes, and a feature that sits at slightly different parameter values across curves is blurred in the mean.
The \emph{elastic} full Procrustes mean additionally quotients out this warping \cite{stoecker2026elasticfullproc}, comparing curves in the square-root-velocity (SRV) framework, under which the reparametrisation group acts by isometries \cite{srivastava2011ElasticCurves,steyer2023ElasticAnalysis}.
We estimate it by alternating covariance smoothing with parametrisation alignment, using the warping alignment for sparse and irregularly observed curves of Steyer et al.~\cite{steyer2023ElasticAnalysis}.
As warping alignment is not rotation-invariant, we first have to estimate an inelastic full Procrustes mean and align each curve rotationally to the current mean.
We then iterate mean estimation, rotational alignment of the observed curves, and warping alignment, giving an iterative estimation scheme.
Each round only reparametrises the data, so training resumes from the current network rather than from a cold start, and the training loss settles at successively lower plateaus until convergence of the alignment.
\begin{algorithm}[htb!]
\caption{Conditional elastic full Procrustes mean estimation}
\label{alg:elastic}
\begin{algorithmic}[1]
\Require
  obs.~$\{(\vec x_i, \vec y_i)\}_{i=1}^{n}$,
  basis $\vec b$,
  Gram $\mat B $,
  reg.~$\rho_t, \{\rho_j\}$
\State $t_{ij} \gets$ arc-length parametrisation of $\vec y_i$ for all $i,j$
\State $\tilde y_{ij} \gets (y_{ij}-\bar y_i)/\norm{y_i-\bar y_i}_w$ for all $i,j$
\Repeat
  \State fit $\vec\theta_0,\{\mat W_j,\vec\varphi_j\}$ by minimising~\eqref{eq:objective} via SGD
  \For{$i = 1,\dots,n$}
    \State $\mat U\mat\Sigma\mat V^{\mathsf H} \gets \mat B^{1/2}\vec\theta(\vec x_i)$
    \State $\hat\mu(t \given \vec x_i) \gets \bigl(\mat B^{-1/2}\vec u_1(\vec x_i)\bigr)\tran \vec b(t)$
    \State $\omega_i \gets \langle \hat\mu(\vec x_i),\, \tilde y_i\rangle$;\quad
           $\tilde y_i \gets (\omega_i/\abs{\omega_i})\,\tilde y_i$
      \Comment{rotational alignment}
    \State $\gamma_i \gets$ warping alignment of $\tilde y_i$ to $\hat\mu(\vec x_i)$
      \Comment{Using Steyer et al.~\cite{steyer2023ElasticAnalysis}}
    \State $\tilde y_i \gets \tilde y_i \circ \gamma_i$;\quad update $t_{ij}$
  \EndFor
\Until{training loss~\eqref{eq:objective} no longer decreases}
\State \Return $\hat\mu(\,\cdot \given \vec x),\ \vec\theta_0,\{\mat W_j,\vec\varphi_j\}$
\end{algorithmic}
\end{algorithm}

\section{Simulation Study}
\label{sec:simulation}

We validate the method on a synthetic dataset with a known conditional mean shape and multimidal covariates (tabular and image).
We compare the full method against two variants: estimation without elastic alignment, and constraining the covariance model to rank $1$.
The latter is of interest as the full Procrustes mean is the first eigenfunction of the estimated covariance surface and therefore, technically, a rank $1$ approximation of the surface could allow mean recovery without modelling redundant variation.
The leading eigenfunction would then be directly given by $\vec\theta(\vec x)$ (up to a normalizing constant) and the resulting mean additive in the covariates by construction.
We plan on investigating this approximation in more detail in further work.

\paragraph{Data.}
Each curve is an irregularly sampled curve consisting of two half circles joined at a cusp.
Their shape is set by two covariates: an observed scalar $x_1 \sim \mathrm{Unif}[0,1]$ controlling the upper radius and a latent $x_2 \sim \mathrm{Unif}[0,1]$ controlling the lower swing.
The model never observes $x_2$ directly, only a $32 \times 32$ grayscale image of an isotropic Gaussian blob whose width grows with $x_2$.
Using an image as a covariate makes it necessary to have at least one deep encoder, given by a convolutional neural network (CNN).
Let $\mu(\cdot \given \vec x)$ be the standardised (centred, unit-norm) conditional mean shape. 
Curve $i$ is sampled at $n_i \sim \mathrm{Unif}\{10, \dots, 50\}$ nodes, the endpoints and $n_i - 2$ interior draws from $\mathrm{Unif}[0,1]$.
It is observed as
\begin{equation}
  y_i(t_{ij}) = \beta_i\, e^{i \theta_i} \Big( \mu(t_{ij} \given \vec x_i) + \sum_{l=1}^{2} \xi_{il}\, \psi_l(t_{ij}) \Big) + \delta_i + \sigma\, \varepsilon_{ij},
  \label{eq:dgp}
\end{equation}
with modes of shape variation $\psi_l(t) = \sin(\pi l t)$ and coefficients $\xi_{il} \sim \mathcal{CN}(0, \lambda_l)$, where $\mathcal{CN}(0,\sigma^2)$ denotes a complex normal with variance $\sigma^2$, $\lambda = (0.10^2, 0.05^2)$, random similarity transformation $\beta_i \sim \mathrm{Unif}[0.5, 1.5]$ (scale), $\theta_i \sim \mathrm{Unif}[0, 2\pi)$ (rotation) and $\delta_i \sim \mathcal{CN}(0, \tau^2)$ (translation, $\tau = 0.5$), and i.i.d.\ isotropic measurement noise $\varepsilon_{ij} \sim \mathcal{CN}(0, 1)$ with $\sigma = 0.02$.
The conditional mean $\mu(\cdot \given \vec x)$ is known in closed form and serves as ground truth.

\begin{figure}[htb!]
  \centering
  \resizebox{\textwidth}{!}{\input{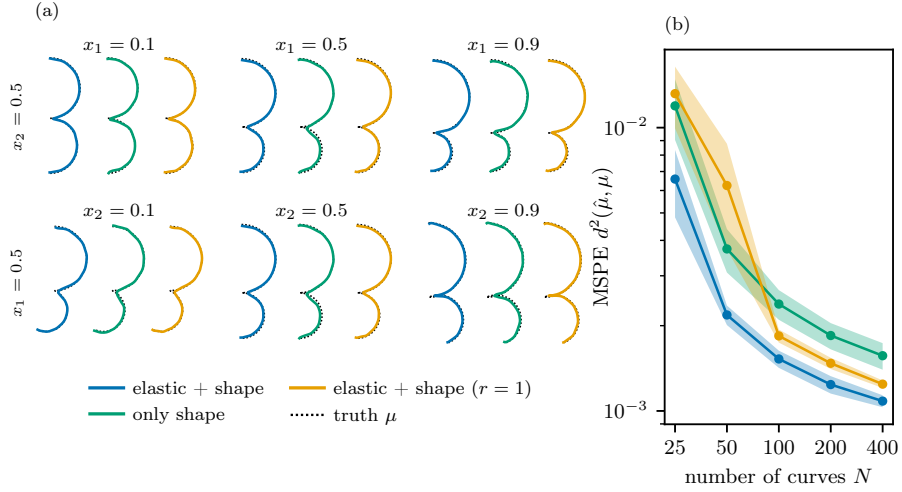}}
  \caption{(a)~Conditional means from a single representative dataset of $N = 400$ curves.
  The top row varies the observed $x_1$ at a fixed sample image ($x_2 = 0.5$) and the bottom row the latent $x_2$ at fixed $x_1 = 0.5$.
  (b)~Recovery error over number of curves $N$ in the dataset, the mean squared prediction error (MSPE) measured using the full-Procrustes distance, after parametrisation alignment, between estimated and true conditional mean over a held-out test set of $100$ observations (mean over $10$ simulations, band $\pm 1$ standard error).}
  \label{fig:simulation_study}
\end{figure}

\paragraph{Model.}
We fit the conditional covariance, with basis $\vec b(t)$ and an additive covariate map $\vec\varphi = [\vec\varphi_1, \vec\varphi_2]$, that is a mix of a deep encoder and penalized b-splines (P-splines) \cite{eilers1996flexible}:
\begin{itemize}
  \item[$\vec b$:] cubic P-spline ($\rho_t = 10^{-3}$), $k = 50$, interior knots uniform on $[0,1]$.
  \item[$\vec\varphi_1$:] cubic P-spline ($\rho_x = 10^{-2}$), $k_x = 10$, interior knots uniform on $[0,1]$.
  \item[$\vec\varphi_2$:] convolutional encoder over the $32 \times 32$ blob image --- three stride-$2$ blocks ($8 \to 16 \to 32$ channels, $3 \times 3$ kernels, batch normalisation, ELU), global average pooling and a linear map.
\end{itemize}
The conditional covariance $\hat C(s,t \given \vec x) = \vec b(s)\tran \mat\Xi(\vec x)\, \vec b(t)$ is factored at rank $r = 3$ (for the mean and the two simulated modes of shape variation $\psi_l$).
We train each fit by stochastic gradient descent (AdamW, cosine-annealing schedule, learning rate $3 \times 10^{-3}$, smoother weight decay $10^{-4}$) on minibatches of $24$ curves for $600$ steps, keeping the parameters at the lowest held-out covariance loss. The elastic alignment  alternates warping with refitting until the held-out loss decreases by less than $1\%$.
We observe that often one elastic alignment step is already enough.
The roughness penalty over $t$ is set to $\rho_t = 10^{-3}$ (via $5$-fold cross-validation of the held-out covariance loss at $N = 400$).
We compare the full method against two variants: dropping the elastic alignment (only shape), and modelling the covariance as rank $1$, sharing all hyperparameters.

\paragraph{Results.}
Figure~\ref{fig:simulation_study} reports the results.
Panel~(a) shows that all three estimators recover the conditional mean shape.
However, using no elastic alignment loses precision at the cusp of the curves, which gets washed out due to parameterisation misalignment.
Panel~(b) gives the recovery MSPE against the number of curves $N$.
The full method is the most accurate at every $N$ and both variants result in consistently higher MSPE.
Despite the slightly higher MSPE, the rank $1$ approximation is visually still very similar to the full method and may serve well as an approximate method, confirming our earlier hypothesis.

\section{Application}
\label{sec:application}

We model the outlines of a slice of the left hippocampus used in Steyer et al.~\cite{steyer2026}, each an open planar curve segmented from a subject's structural MRI scan along the hippocampus' longest axis.
The outlines are prealigned and scaled, but only approximately, so estimating shape rather than  relying on the raw coordinates remains of interest.
For every subject we observe the tabular covariates age, sex, \emph{Alzheimer's disease} (AD) diagnosis and the APOE-$\varepsilon$4 allele dosage, i.e.\ the number of $\varepsilon$4 alleles a subject carries ($0$, $1$ or $2$: non-carrier, heterozygous or homozygous), as well as an sMRI of the subject's whole brain (pre-processing details are given in the Appendix).

\begin{figure}[!htb]
  \centering
  \subfloat[Model with \emph{Age} range given by the 5th--95th percentile of the age distribution, \emph{Group} contrasting \emph{Alzheimer's disease} (AD) with cognitively normal (CN), \emph{APOE4} the $\varepsilon$4-allele dosage, and \emph{Sex} contrasting female (F) and male (M).]{%
    \resizebox{\textwidth}{!}{\input{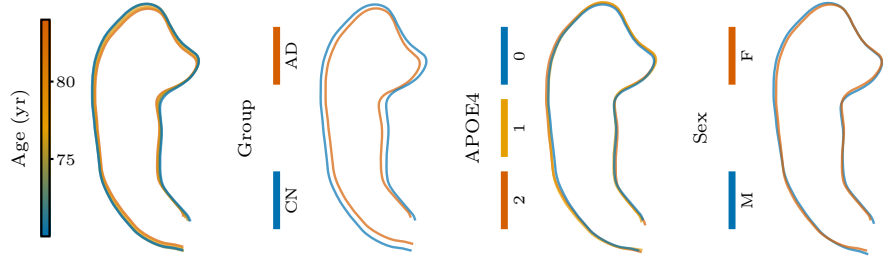}}%
    \label{fig:adni_effects_add}}\\
    \subfloat[Model with an added AD\,$\times$\,APOE-$\varepsilon$4 interaction, coded as $1$ for AD subjects carrying at least one $\varepsilon$4 allele (e4+) and $0$ otherwise (e4-).]{%
    \resizebox{\textwidth}{!}{\input{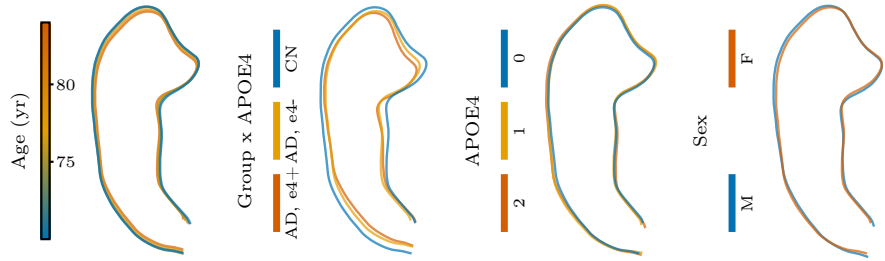}}%
    \label{fig:adni_effects_int}}
  \caption{Partial covariate effects on the left hippocampus outline ($n = 239$). Each panel varies one covariate with the others held at the reference subject (median age, female, cognitively normal, no APOE-$\varepsilon$4) and shows the conditional size-and-shape mean.
  }
  \label{fig:adni_effects}
\end{figure}

\paragraph{ADNI Dataset.}
Data used in the preparation of this article were obtained from the Alzheimer's Disease Neuroimaging Initiative (ADNI) database.
The ADNI was launched in
2003 as a public-private partnership, led by Principal Investigator Michael W. Weiner,
MD. The primary goal of ADNI has been to test whether serial magnetic resonance imaging
(MRI), positron emission tomography (PET), other biological markers, and clinical and
neuropsychological assessment can be combined to measure the progression of mild
cognitive impairment (MCI) and early Alzheimer's disease (AD). For up-to-date information,
see \url{www.adni-info.org}.

\paragraph{Modelling shape and size.}
Age and AD status are known to reduce hippocampal volume \cite{henneman2009},
but the conditional full Procrustes mean $\mu(t \given \vec x)$ is unit-norm and scale-invariant by construction, so it carries no size information and cannot recover such effects.
We therefore model size explicitly and report \emph{size-and-shape} means.
Each curve is already scaled by its weighted centroid size, $S_i = \norm{y_i - \bar y_i}_w = ( \textstyle\sum_j w_{ij}\,\abs{y_{ij}-\bar y_i}^2)^{1/2}$, which we retain as the response of a separate log-size regression of $\log S_i$ on the same covariates, giving a conditional size $S(\vec x)$.
The conditional size-and-shape mean is the unit-norm elastic full Procrustes mean scaled by the predicted size,
\begin{equation}
  \hat m(t \given \vec x) = \hat S(\vec x)\,\hat\mu(t \given \vec x).
\end{equation}
Keeping scale out of the covariance leaves the shape estimand scale-invariant, while $S(\vec x)$ lets each covariate act on size.
The results are reported in Figure~\ref{fig:adni_effects}.

\paragraph{Model.}
We use the same encoder architecture for the shape and size models but fit them separately.
The curve basis $\vec b(t)$ uses cubic P-splines, $k = 128$, $age$ enters through a penalised cubic b-spline ($k = 10$), sex and AD enter linearly as binary indicators, and the APOE-$\varepsilon$4 dosage is treated as categorical ($0 =$ non-carrier, $1 = $ heterozygous, $2=$ homozygous).
We set the rank of the conditional covariance to $r = 4$ after inspecting the scree plot of a higher-rank fit, depicted in the Appendix.
We train using stochastic gradient descent (AdamW, cosine-annealing schedule, learning rate $3 \times 10^{-3}$, smoother weight decay $10^{-4}$), $64$ curves per step, with early stopping on a held-out validation set.
The roughness penalty $\rho_t = 1$ and the scale-model weight decay $10^{-2}$ are selected by $5$-fold cross-validation.

\paragraph{Results.}
Figure~\ref{fig:adni_effects_add} reports the partial effect of each covariate on the conditional mean outline, with the others held at the reference subject.
We find that increasing $age$ and having $AD$ both have a shrinking effect on the hippocampus, which is consistent with the earlier analysis of \cite{steyer2026} and the literature on hippocampal shape \cite{henneman2009}.
$sex$ has no effect in the dataset, likely because the hippocampus outlines were extracted from aligned MRI which were already volume matched.
We also modelled the effect of APOE-$\varepsilon$4 dosage, first without accounting for interaction with $AD$, and find only tiny differences across categories.
However, when estimating a model with an AD\,$\times$\,APOE-$\varepsilon$4 interaction we find that AD is associated with somewhat greater hippocampal shrinkage in $\varepsilon$4 carriers than in non-carriers, which is also reported in the literature \cite{pievani2011APOE4}.

\paragraph{Outlook.}
We limit our main analysis to tabular covariates.
However, to give a first impression of the capabilities of a multimodal model, we provide a sensitivity analysis in the Appendix, where we control for per-subject brain structure by including a subjects' sMRI scan (with the hippocampus blocked out) in the model.
We then compare difference in the estimated AD effect across models.
In the future, we would like to more thoroughly investigate the effects of high-dimensional covariates, such as genomics, and possibly model the full hippocampal volume by, for example, modelling it jointly as an array of slices along the hippocampus.

\section{Conclusion}
\label{sec:outlook}
We have shown that the conditional full Procrustes mean can be estimated via a deep conditional covariance smoother that admits multimodal covariates through learned encoders.
Iterative mean estimation and parametrisation alignment allow us to estimate elastic means, recovering crisp mean curves in a simulation study with multimodal covariates.
We provided a first application to hippocampal size-and-shape regression that provided results consistent with the literature, which we want to extend going forward.
Methodologically, several directions remain open.
The penalty parameters $\rho_t$ and $\{\rho_j\}_j$ are currently selected by cross-validation, whereas penalised-spline smoothing parameters are chosen in a more principled way by restricted maximum likelihood \cite{wood2011fast}.
This approach does not extend easily to models with neural encoders, and finding a comparable smoothness criterion is an interesting open question.
Our elastic mean treats parametrisation alignment as an outer alternating loop around the smoother and uses an established R package for estimation.
Implementing it in a deep learning framework would greatly speed up computations.
Beyond the leading eigenfunction, the fitted covariance already estimates modes of shape variation, which might be used for inference and uncertainty estimation in the future.
The present model treats open curves, and an extension to closed curves would require a periodic basis together with a closure constraint, further broadening applicability of our method.
Finally, the Hermitian characterisation exploits the identification $\mathbb{R}^2 \cong \mathbb{C}$, so an extension to three-dimensional shapes would be of interest but would lose this structure and would need a different algebraic treatment.
\begin{credits}
\subsubsection{\ackname}
Funded by the Deutsche Forschungsgemeinschaft (DFG, German Research Foundation) – project number 459422098.
Data collection and sharing for the Alzheimer's Disease Neuroimaging Initiative (ADNI) is
funded by the National Institute on Aging (National Institutes of Health Grant U19
AG024904). The grantee organization is the Northern California Institute for Research and
Education.
In the past, ADNI has also received funding from the National Institute of Biomedical Imaging and Bioengineering, the Canadian Institutes of Health Research, and private sector contributions through the Foundation for the National Institutes of Health (FNIH) including generous contributions from the following: AbbVie, Alzheimer's Association; Alzheimer's Drug Discovery Foundation; Araclon Biotech; BioClinica, Inc.; Biogen; Bristol-Myers Squibb Company; CereSpir, Inc.; Cogstate; Eisai Inc.; Elan Pharmaceuticals, Inc.; Eli Lilly and Company; EuroImmun; F.~Hoffmann-La Roche Ltd and its affiliated company Genentech, Inc.; Fujirebio; GE Healthcare; IXICO Ltd.; Janssen Alzheimer Immunotherapy Research \& Development, LLC.; Johnson \& Johnson Pharmaceutical Research \& Development LLC.; Lumosity; Lundbeck; Merck \& Co., Inc.; Meso Scale Diagnostics, LLC.; NeuroRx Research; Neurotrack Technologies; Novartis Pharmaceuticals Corporation; Pfizer Inc.; Piramal Imaging; Servier; Takeda Pharmaceutical Company; and Transition Therapeutics.

\subsubsection{\discintname}
The authors have no competing interests to declare that are relevant to the content of this article.
\end{credits}

\appendix

\section*{Appendix}

\subsection*{The Closed-Form Full Procrustes Distance}
\label{app:fpdist}
Fix $\mu$ with $\norm{\mu} = 1$ and a realisation $\tilde y$ with $\norm{\tilde y} = 1$. Expanding the Hermitian norm and using $\langle \tilde y, \mu\rangle = \overline{\langle \mu, \tilde y\rangle}$,
\begin{equation*}
  \norm{\mu - \omega \tilde y}^2
  = \norm{\mu}^2 - \overline{\omega}\,\langle \mu, \tilde y\rangle - \omega\,\langle \tilde y, \mu\rangle + \abs{\omega}^2 \norm{\tilde y}^2
  = 1 + \abs{\omega}^2 - 2\,\operatorname{Re}\!\big(\omega\,\langle \tilde y, \mu\rangle\big).
\end{equation*}
Completing the square in $\omega$ gives $\norm{\mu - \omega \tilde y}^2 = 1 - \abs{\langle \mu, \tilde y\rangle}^2 + \abs{\omega - \langle \mu, \tilde y\rangle}^2$, minimised at $\omega^\star = \langle \mu, \tilde y\rangle$ with value
\begin{equation*}
  \min_{\omega \in \C} \norm{\mu - \omega \tilde y}^2 = 1 - \langle \mu, \tilde y\rangle \langle \tilde y, \mu\rangle .
\end{equation*}

\subsection*{Computational Cost}
\label{app:compute}
All experiments ran on an NVIDIA A100 GPU. 
The simulation study in Section~\ref{sec:simulation} runs a sweep over the five sample sizes $N \in \{25, 50, 100, 200, 400\}$, with ten repetitions and three method variants each, and completes in about $30$ minutes.
The fitting time is dominated by the iterative elastic alignment using the R-package \texttt{elasdics} \cite{steyer2023ElasticAnalysis} rather than by the covariance smoothing step.
For future work, we propose implementing the parametrisation alignment step directly in a deep learning framework, as it is currently CPU bound.

\subsection*{Further Details on the Application}
\label{app:adni-hp}
\begin{figure}[!htb]
  \centering
  \includegraphics[trim=0 0 186 0, clip, height=110pt]{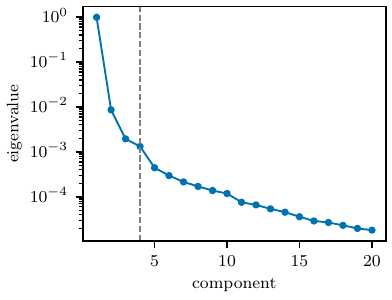}\hfill
  \includegraphics[height=110pt]{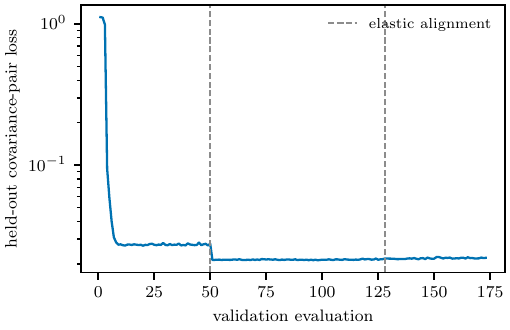}
  \caption{Fitting diagnostics for the model using only tabular covariates.
  \emph{Left}: eigenvalues on a log scale from eigendecomposition of the covariance surface (fit at $r = 32$).
  \emph{Right}: held-out covariance loss over training, where the dashed lines mark elastic-alignment steps.}
  \label{fig:adni-scree}
  \label{fig:elastic_staircase}
\end{figure}

The covariance rank was determined heuristically by looking at a scree plot of a higher-rank fit. 
Figure~\ref{fig:elastic_staircase} tracks the held-out covariance loss during fitting and demonstrates the effect of elastic alignment, which lowers the convergence floor for the held-out loss.
Both figures correspond to the model using tabular covariates and no interaction effect.
The sMRI scans were affine-registered to the MNI152 template, obtained from TemplateFlow~\cite{ciric2022TemplateFlow}, using ANTs~\cite{avants2011ReproducibleEvaluation,tustison2021ANTsXEcosystem}, and skull-stripped with HD-BET~\cite{isensee2019BrainExtraction}.
We then blanked both hippocampi from each scan by placing ellipsoids at the centroids of the hippocampi, oriented to their principal axes, and blanking the voxels there.
The size of the ellipsoids was fixed to the largest per-axis half-extent of any hippocampus plus a small margin.

\subsection*{Sensitivity of the AD Effect to the Surrounding Brain}
\label{app:adni-sensitivity}

\begin{figure}[!htb]
  \centering
  \includegraphics[width=\textwidth]{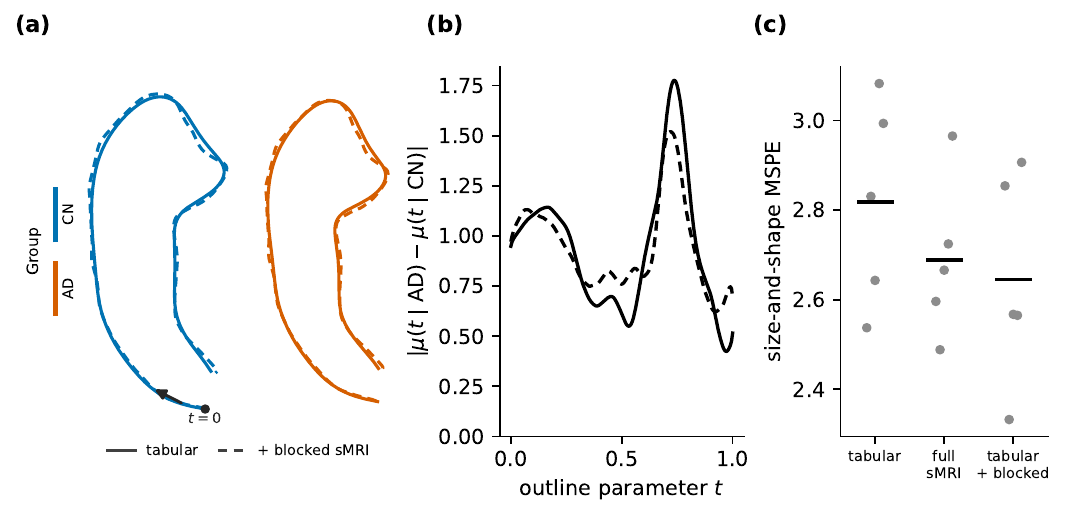}
  \caption{Sensitivity of the estimated AD effect to adjusting for the surrounding brain, comparing a model that uses the tabular covariates alone (solid) against one that additionally conditions on the hippocampal-blocked sMRI scan (dashed).
  \emph{(a)}~Predicted control (blue) and AD (orange) size-and-shape outlines, marginalised over the held-out test set.
  \emph{(b)}~Magnitude of the AD effect along the outline.
  \emph{(c)}~Cross-validated size-and-shape mean squared prediction error of three models with a bar marking the mean.}
  \label{fig:adni_sensitivity}
\end{figure}

To check whether the estimated AD effect is sensitive to the inclusion of the sMRI scan in the model, we compare the effect estimated from the tabular covariates alone against the effect when the model additionally conditions on the hippocampal-blocked sMRI scan.
Both share the same fitting recipe, and each AD effect is marginalised over the held-out test set. 
As Figure~\ref{fig:adni_sensitivity} shows, adding the scan seems to make the estimated outlines slightly sharper, but this is only a mild change in some parts of the shape.
Adding sMRI scans or even predicting the hippocampal shape using a full sMRI scan leads to slightly better mean shape recovery on held-out datasets.
In the future it would be interesting to calculate the statistical significance of these differences using, perhaps, the full information of the covariance surface, which encodes shape variation across the sample.

\bibliographystyle{splncs04}
\bibliography{references}

\end{document}